\documentclass[a4paper,fleqn]{cas-sc}

\usepackage[numbers]{natbib}
\def\tsc#1{\csdef{#1}{\textsc{\lowercase{#1}}\xspace}}
\tsc{WGM}
\tsc{QE}
\newcommand{\researchQuestion}[2]{$#1 _ #2$}

\begin{document}
\let\WriteBookmarks\relax
\def\floatpagepagefraction{1}
\def\textpagefraction{.001}

\shorttitle{SelfIE: Self-Initiated Explorable Instructions}    
\shortauthors{Hyeongcheol Kim}  

\title [mode = title]{SelfIE: Self-Initiated Explorable
Instructions Towards Enhanced User Experience}  

%$\author[<aff no>]{<author name>}[<options>] ---------------------1st
\author[1]{Hyeongcheol Kim}[type=editor,
       style=korean,
       bioid=1,
  %     prefix=,
       orcid=0000-0003-4327-2148,
       facebook=danielKim0911,
       twitter=bluesky0911,
       linkedin=hyeongcheol-kim-76043844]

%\cormark[<corr mark no>]
\cormark[1]

%\fnmark[<footnote mark no>]

\ead{hckim0911@gmail.com}

\ead[url]{https://hckim.net}

\credit{Conceptualization of this study, Methodology, Software, Validation, Formal analysis, Investigation, Data Curation, Writing - Original Draft, Writing - Review \& Editing, Visualization}

\affiliation[1]{organization={Synteraction Lab, National University of Singapore},
            addressline={School of Computing, COM1, 13, Computing Dr}, 
            city={Singapore},
%          citysep={}, % Uncomment if no comma needed between city and postcode
            postcode={117417}, 
            country={Singapore}}

%\author[<aff no>]{<author name>}[<options>] ---------------------2nd
\author[2]{Katherine Fennedy}[type=editor,
       style=singaporean]
 %      auid=000,
 %      bioid=1,
  %     prefix=,
%       orcid=0000-0002-9643-2072,

% Footnote of the second author

% Email id of the second author
\ead{katherine.fennedy@gmail.com}


\credit{Methodology, Validation, Writing - Review \& Editing}

\affiliation[2]{organization={DQ LAB Pte Ltd},
            addressline={20a Tanjong Pagar Road}, 
            city={Singapore},
%          citysep={}, % Uncomment if no comma needed between city and postcode
            postcode={088443}, 
            %state={},
            country={Singapore}}

% \cortext[1]{Corresponding author}


%\author[<aff no>]{<author name>}[<options>] ---------------------3rd
\author[3]{Georgia Zhang}[type=editor,
    style=singaporean]
%    gplus=<gplusorean>,
  %     prefix=Miss,
 %      orcid=0000-0003-1718-8022]
 %         facebook=<facebook id>,
 %        linkedin=<linkedin id>,
  %      gplus=<gplus id>]

% Footnote of the second author
%\fnmark[3]

% Email id of the second author
\ead{jiayi.zhang@u.nus.edu}

%\ead[url]{https://jiajiazhang221.wixsite.com/my-site}

\credit{Data Curation, Investigation}

\affiliation[3]{organization={National University of Singapore},
            addressline={3 Research Link}, 
            city={Singapore},
%          citysep={}, % Uncomment if no comma needed between city and postcode
            postcode={117602}, 
            state={N/A},
            country={Singapore}}

% \cortext[1]{Corresponding author}


%\author[<aff no>]{<author name>}[<options>]---------------------4th
\author[4]{Can Liu}[type=editor,
       style=chinese]
%       prefix=,

% Footnote of the second author

% Email id of the second author
\ead{canliu@cityu.edu.hk}

%\ead{}

\credit{Methodology, Validation}

\affiliation[4]{organization={School of Creative Media, City University of Hong Kong},
            addressline={Run Run Shaw Creative Media Centre, Level 7, 18 Tat Hong Avenue, Kowloon Tong}, 
            city={Hong Kong},
%          citysep={}, % Uncomment if no comma needed between city and postcode
            country={China}}
            
% \cortext[1]{Corresponding author}


%\author[<aff no>]{<author name>}[<options>] ---------------------5th
\author[5]{Shengdong Zhao}[type=editor,
       style=canadian]
%       prefix=,

% Footnote of the second author

% Email id of the second author
\ead{shengdong.zhao@cityu.edu.hk}

% \ead{}

\credit{Supervision, Project administration, Writing - Review \& Editing, Funding acquisition}

\affiliation[5]{organization={School of Creative Media and Computer Science Department, City University of Hong Kong},
            addressline={83 Tat Chee Avenue, Kowloon}, 
            city={Hong Kong},
%          citysep={}, % Uncomment if no comma needed between city and postcode
            country={China}}

% \cortext[1]{Corresponding author}




%\tnotetext[<tnote number>]{<tnote text>} 


\begin{abstract}
Given the widespread use of procedural instructions with non-linear access (situational information retrieval), there has been a proposal to accommodate both linear and non-linear usage in instructional design. However, it has received inadequate scholarly attention, leading to limited exploration. This paper introduces \textbf{Self}-\textbf{I}nitiated \textbf{E}xplorable (SelfIE) instructions, a new design concept aiming at enabling users to navigate instructions flexibly by blending linear and non-linear access according to individual needs and situations during tasks. Using a Wizard-of-Oz protocol, we initially embodied SelfIE instructions within a toy-block assembly context and compared it with baseline instructions offering linear-only access (N=21). Results show a 71\% increase in user preferences due to its ease of reflecting individual differences, empirically supporting the prior proposal. Besides, our observations identify three strategies for flexible access and suggest the potential of enhancing the user experience by considering cognitive processes and implementing flexible access in a wearable configuration. Following the design phase, we translated the WoZ-based design embodiment as working prototypes on the tablet and OHMD to assess usability and compare user experience between the two configurations (N=8). Our data yields valuable insights into managing the trade-offs between the two configurations, thereby facilitating more effective flexible access development.
\end{abstract}


\begin{highlights}
\item We propose a new instructional design concept: \textbf{Self}-\textbf{I}nitiated \textbf{E}xplorable instructions.
\item SelfIE instructions provide users with easy access to situationally necessary information in instructional materials during tasks, seamlessly integrated with linear access.
\item SelfIE assembly instructions, based on this concept, significantly improve user preferences compared to baseline instructions that provide only linear access due to its ease of reflection in individual differences.
\item Incorporating cognitive process support and wearable technologies in SelfIE assembly instructions suggests the potential for further enhancing user experience.
\end{highlights}

% Keywords

\maketitle

% Main text
\section{Introduction}
\label{SELFIE:INTRODUCTION}
While it is easily assumed that people will read and follow procedural instructions step-by-step from beginning to end, a body of research on how people use instructions \cite{szlichcinski1979telling,wright1982,schriver1997dynamics,szlichcinski1979telling,carroll1990nurnbergfunnel,rettig1991nobody,redish1998minimalism} suggests the necessity of revisiting this assumption. Studies dating back to the 1980s indicate that people often approach instructions in a non-linear fashion (e.g., scanning first and referring to necessary parts only \cite{ganier2004factors}) because such usages can reduce cognitive demands from reading instructions, improve efficiency in specific situations (e.g., for simple tasks \cite{wright1982} or in having foreknowledge \cite{ganier2004factors,wright1982}), and align better with their learning or task strategies \cite{ganier2004factors, carroll1990nurnbergfunnel}. These non-linear approaches appear across various tasks \cite{schriver1997dynamics, wright1982,szlichcinski1979telling, carroll1990nurnbergfunnel}, even if the instructions are originally designed for users' step-by-step use \cite{rettig1991nobody}.

Given this widely observed phenomenon, prior researchers have proposed the need to accommodate linear and non-linear instructional usage in the design of instructions \cite{eiriksdottir2011procedural,ganier2004factors}. However, such a proposal has garnered limited scholarly attention thus far, leading to a dearth of exploration into its effective implementation, usages, and empirical evaluation of potential benefits for users.

This paper introduces \textbf{Self}-\textbf{I}nitiated \textbf{E}xplorable (SelfIE) instructions, a new design concept that aims at enabling users to navigate instructions flexibly by blending linear (i.e., moving forwards or backwards a step to proceed in the task) and non-linear (i.e., jumping to a certain step to retrieve necessary information within instructions during the task) access. With such flexible access, users are expected to tailor their instructional use to individual needs and situations in their task completion.

To practically explore the concept, we initially embodied it in a toy-block assembly context (i.e., SelfIE assembly instructions). Considering prior guidelines for assembly instruction designs and aids, we designed assemblers' flexible access to instructional materials using a show-and-ask metaphor, anchored in a requester-responder interaction model. Thereby, using SelfIE assembly instructions, the assembler can navigate instructional materials in a multi-modal manner by verbalising requests for assistance (i.e., by uttering ``next'' or ``previous'' for linear access), if necessary, by displaying any target component(s) (i.e., by uttering ``this one'' with showing assembly block(s) for non-linear access). Following this, we conducted two rounds of design iterations with the initial design, using a Wizard-of-Oz (WoZ) protocol \cite{dow2005wizard} to ensure its effectiveness. As a result of this, we made two design adjustments: 1) restricting non-linear access to be used only with a single block (instead of a chunk) to clarify its operation mechanism as 1-to-1 mapping between the displayed block and its corresponding instructional step; 2) allowing the simulator's heuristic decision-making for remaining ambiguous scenarios where a single block could correspond to multiple steps, along with the assembler's unidentifiable status.

After the design embodiment phase, we investigated the usability and user experience of SelfIE assembly instructions by using the same WoZ protocol compared to baseline instructions offering linear-only access (N=21). Participants assembled two models, deliberately adjusted to a similar level of complexity, using the two types of instructions in an open-formed assembly task. Results showed a 71\% increase in user preferences due to its ease of reflecting individual differences in instructional use, empirically supporting the prior proposal. Furthermore, our video analysis identified three particular flexible access strategies employed by assemblers: selective skipping, debugging, and block-scanning, providing an understanding of how they customised their instructional uses with flexible access to suit their needs and situations during assembly and why the ability led to enhanced preferences in using instructions. Apart from these findings, our observational and qualitative data suggested the potential of incorporating cognitive process support in using flexible access and wearable technologies for a further enhanced user experience.

Additionally, we transformed our WoZ-based design work into a working prototype to explore its feasibility and usability under real-world constraints. Given the recognised potential of wearable technologies (e.g., smart glasses) in our prior study, our prototype incorporated both a tablet and an optical head-mounted display (OHMD). We then qualitatively evaluated usability and user experience by applying our prior experimental protocol to these two prototype forms (N=8). In interviews, participants strongly supported the potential of our design approach despite the limited usability arising from technical imperfections in its implementation. Besides, their comparative feedback revealed the trade-offs between the tabletop and wearable configuration. Wearable sensors facilitated the wearer’s physical interactions to trigger non-linear access to instructions during assembly. However, the wearable display of instructional materials, floating in front of the eyes and moving with the wearer's head movements, presented challenges in effectively locating and reading the materials, alongside the display’s low resolution. Considering these trade-offs, we discussed the potential for a hybrid approach that combines the stability of the tabletop display (tablet) with the mobility of the wearable display (OHMD). This strategy aims to maximally leverage the strengths of both configurations to enhance user experiences further using instructions with flexible access.

\vspace{0.6cm}

This paper's contributions are threefold:

\vspace{0.4cm}
\begin{enumerate}
    \item Introduction of a new instructional design concept: SelfIE instructions, aiming at providing users with flexible access to instructional materials during a task;
    \item Initial design embodiment of SelfIE instructions within a toy-block assembly context and investigation of its usages and experiences via WoZ protocol, prompting further discussions on enhancing them;
    \item Prototyping on the tabletop and wearable configuration, along with assessing the usability and the potential of wearable configuration for further enhanced user experiences.
\end{enumerate}
\section{Background}
\label{SELFIE:BACKGROUND}
\subsection{Linear Instruction Usage and User Reluctance}
Procedural instructions describe step-by-step actions to complete a procedural task \cite{guthrie1991, farkas1999logical}. With this nature, many researchers and designers can easily assume that people will willingly read their instructions before engaging in a procedural task and follow them linearly during the task, from beginning to end. However, contrary to this assumption, a body of research on how people use procedural instructions indicates that many people are reluctant to use instructions linearly due to three primary reasons: 1) instructions themselves can sometimes be too long and difficult to follow \cite{szlichcinski1979telling, redish1998minimalism}; 2) if the task is too simple \cite{wright1982} or the quality of the instructions is too low \cite{wieringa1998procedure}, instructions can be perceived as useless or 3) if they desire to carry out the task by their ability \cite{carroll1990nurnbergfunnel}, instructions can be referenced at needs only.

\subsection{Actual Instruction Usage}
Prior research suggests that many use instructions in a different order with non-linear access according to their needs and situations during tasks. For example, novices or those with a cautious approach tend to read instructions from start to finish and follow them step-by-step during a task (i.e., instruction-first approach) \cite{ganier2004factors}. In contrast, experienced or adventurous people tend to use instructions as needed while figuring out what they can do independently (i.e., task-first approach) \cite{carroll1990nurnbergfunnel}. Besides the two contrasting groups of people, some can scan instructions first and then refer back to necessary parts for clarification (i.e., selective reference) \cite{schriver1997dynamics}; or while others can only refer to instructions when unsure what to do (i.e., demand-based reference) \cite{wright1982}.

These flexible instructional usages, i.e., non-linear access combined with linear access for necessary information retrieval within instructions, are prevalent across various procedural systems, from household appliances \cite{schriver1997dynamics} to computer software \cite{carroll1990nurnbergfunnel}. This approach persists even when instructions are designed for sequential following, and despite some benefits of adhering strictly to linear instructions for task completion \cite{carroll1985exploring, carroll1990nurnbergfunnel}. Recognising this, previous researchers have advocated for the inclusion of non-linear instructional strategies in instructional design \cite{ganier2004factors, eiriksdottir2011procedural}. Ganier \cite{ganier2004factors} suggests that such design considerations align with natural user strategies, potentially improving usability and enhancing overall user experience. Furthermore, Carroll \cite{carroll1990nurnbergfunnel} notes that incorporating these considerations could increase user motivation and engagement, fostering deeper exploration and learning.

\subsection{Three Difficulties in Using Instructions}
Ganier \cite{ganier2004factors} categorised the challenges users encounter with instructional materials into three main areas: 1) understanding the materials, 2) applying this understanding to specific situations, and 3) locating necessary information within the materials as needed. These categories offers a framework for organising existing research aimed at enhancing usability and user experience based on the specific challenge that they addressed.

\subsubsection*{High Fidelity of Instructions}
From the user's perspective, high-quality instructions can minimise cognitive load for reading and understanding, threby enhancing usability and experience. This effect has been demonstrated by numerous studies (e.g., \cite{van2014comparison,chirumalla2015influence,guan2006effects}), suggesting that involving higher fidelity instructional materials (e.g., images or videos) lowers the required cognitive effort for comprehension of the instructional materials compared to lower fidelity ones (e.g., text-only instructions). Based on this effect, various high-fidelity instructions have been proposed and explored, such as mixing images and videos for easier comprehension \cite{chi2012mixt}, employing dimensional (3D) tracking capabilities for complex assembly tasks \cite{gupta2012duplotrack}, or implementing multimedia instructions for efficient learning \cite{michas2000learning}. User evaluations of such high fidelity instructions have consistently indicated improved usability and overall user experience.

\subsubsection*{Easy Application of Instructions}
Several works have improved usability and user experience by adapting instructional presentations to users' dynamic states during tasks and their perspectives. For instance, Smart Makerspace \cite{knibbe2015smart} aids users by providing spatially and contextually relevant information, such as tool locations, tailored to the specific step of the instruction they are on. Similarly, Pause-and-Play \cite{pongnumkul2011pause} enhances instruction following by automatically pausing and resuming tutorial videos to match the user's pace. Furthermore, Evertutor \cite{wang2014evertutor} automatically creates interactive, guided tutorials on smartphones based on users’ actual actions. In addition to these examples, Funk et al. \cite{funk2016interactive, funk2018teach} have shown how in-situ projected instructions can simplify users’ application of instructions. Likewise, Yang et al. \cite{yang2020comparing} investigated the effects of different instructional presentations (e.g., augmented versus paper-based instructions) on users’ comprehension and application of instructions. Moreover, with recent advances in Optical Head-Mounted Display (OHMD) technology, numerous studies have explored how to effectively utilise the device’s affordance to enhance users' application of instructions. These studies have shown that OHMDs can benefit users from applying instructions by rendering in-view pictorial instructions \cite{blattgerste2018situ} and providing timely, in-situ information for procedural tasks, such as path navigation \cite{roy2017followmylead} and machine manipulation \cite{huang2021adaptutar, leelasawassuk2017automated}.

\subsubsection*{Facilitation of Information Retrieval Within Instructions}
Compared to the aforementioned works, only a limited number of studies have focused on facilitating information retrieval within instructions, specifically aiming to enhance usability and user experience. Regarding the challenge, one conventional solution involves organising instruction contents with coloured or bold headings in a modular manner to facilitate the user's information scanning \cite{hartley1995chapter}. Another typical solution is to append a content index at the end of the instructions to facilitate the user's non-linear references during the task \cite{ganier1999traitement}. To date, these solutions have appeared in many instructional designs, regardless of the type of instructions (i.e., analogue or digital), although the solutions come with another cognitive effort for reading the headings or the index and looking for the exact necessary information from instructions. In the contemporary context, we only identified two relevant works, VideoWhiz \cite{nawhal2019videowhiz} and CodeTube \cite{ponzanelli2016codetube}. VideoWhiz is a new video interface for recipe videos motivated by users' non-linear viewing habits, providing a hierarchical overview of recipe steps from the video for the user's quick overview grabbing. Its interactive UI components also enable users to enter keywords or set filtering options to facilitate searching and watching a necessary part of the recipe video alongside conventional linear video consumption. CodeTube \cite{ponzanelli2016codetube} is a web-based recommender system that analyses the contents of tutorial videos, providing cohesive and self-contained video fragments in response to a given query, along with links to relevant Stack Overflow discussions to the query. Both works focused evaluating their implemented design features and user acceptance of them, and results suggested that such interface designs can elicit a more favourable response from users compared to the interface lacking the ease.
\section{Research Gap}
\label{SELFIE:RESEARCH_GAP}
Our literature review suggests that people often use instructions with non-linear access (i.e., situational information retrieval within instructions) to meet their needs or situations during tasks. Since this flexible instructional usage is widely observed, prior researchers advocated the need to accommodate both linear and non-linear usage in instructional design, with expectation of enhanced usability and user experience when using instructions \cite{ganier2004factors,eiriksdottir2011procedural}. 

Despite its potential, the proposal has received little scholarly attention, as evidenced by the limited contemporary research on facilitating information retrieval within instructions and its impact on users' instructional usages and experiences. Specifically, given the fact that the limited studies, such as  VideoWhiz \cite{nawhal2019videowhiz} and CodeTube \cite{ponzanelli2016codetube}, have predominantly focused on evaluating their design features within the context of developing advanced video interfaces, there is a lack of exploration into the effective implementation of the proposal, its usage patterns, and potential user benefits across various procedural tasks.
\section{Concept Design: Self-Initiated Explorable (SelfIE) Instructions}
\label{SELFIE:CONCEPT_DESIGN}
The research gap identified in Chapter \ref{SELFIE:RESEARCH_GAP} prompted us to devise a new instructional design concept, namely \textbf{Self}-\textbf{I}nitiated \textbf{E}xplorable (SelfIE) instructions. This concept aims at embracing individuals' non-linear approaches within the context of instructional design through enabling users' information retrieval within instructions as easy as their linear access to them. Thereby, users are expected to flexibly navigate instructional materials by blending their non-linear access with linear access. This aspect notably distinguishes our approach from conventional linear-usage-oriented instructional designs or conventional solutions (e.g., appending the index at the end of instructions). Below, we delineate the core design factors of SelfIE instructions and its envisioned usage:

\begin{enumerate}
    \item \textit{\textbf{Three access operations}}: SelfIE instructions enable users' navigation of instructional materials through three mental operations: moving forwards and backwards a step along a series of steps \textit{plus} jumping to a specific step that provides situationally necessary information.
    \item \textit{\textbf{Balancing linear and non-linear access}}: The first two operations require a design and implementation akin to conventional linear access to instructional materials; in contrast, the last operation requires a design and implementation that enables users to both retrieve and access a situationally necessary instructional part via comparable physical and psychological effort to the linear access operations (i.e., achieving in-situ information retrieval within instructional materials in a single mental transaction, akin to linear access operations).
    \item \textit{\textbf{Seamless transition}}: The design and implementation of the three operations require ensuring that users do not experience additional cognitive load during transitions between them (i.e., functional effectiveness of the situational non-linear access operation to confirm that users do not encounter any erroneous behaviours after performing it, akin to the linear access operations).
\end{enumerate}

Prior research suggests that individuals' varied instructional usages originate from their distinct learning strategies or situations \cite{eiriksdottir2011procedural}, and there is a recognised link between these diverse instructional usages and non-linear access to materials \cite{ganier2004factors}. Building on these insights, we envision that users could navigate SelfIE instructions flexibly, tailoring their approach to individual circumstances (Self-Initiated) by integrating three access operations during tasks (Explorable). For instance, individuals who prefer selective referencing \cite{schriver1997dynamics} might initially browse the instructional materials using primarily forward or backward linear access. They could then seamlessly combine  non-linear access with linear access to delve into specific or adjacent steps when complexities arise or details need to be revisited. Conversely, experienced or adventurous users \cite{carroll1990nurnbergfunnel} might rarely use linear access during tasks, as they generally consult the materials less frequently. Instead, they might occasionally utilise non-linear access combined with linear access to address challenges that they cannot resolve independently.
\section{Design Embodiment: SelfIE Assembly Instructions}
\label{SELFIE:DESIGN_EMBODIMENT}
To explore SelfIE instructions practically, we chose a toy-block assembly task (e.g., LEGO®) for its inherent flexibility and adaptability. This task aligns with SelfIE's design to support users’ flexible navigation based on individual needs and task contexts, allowing for variable task complexity and accommodating different expertise levels (novice vs expert) and performance styles (instruction-based vs task-based). Unlike tasks requiring strict sequential steps, such as knotting \cite{meilach1971macrame} or origami, toy-block assembly permits exploration and adaptation. Drawing from Eiriksdottir's insights (2008) \cite{eiriksdottir2008people}, this selection also aims to enrich our understanding of usability and user experiences with SelfIE instructions, with the task's enjoyable nature potentially broadening users' engagement purposes, from recreation to efficiency-driven completion.

\subsection{Initial Design}
\begin{figure}[hbt!]
    \centering
    \includegraphics[width=0.9\textwidth]{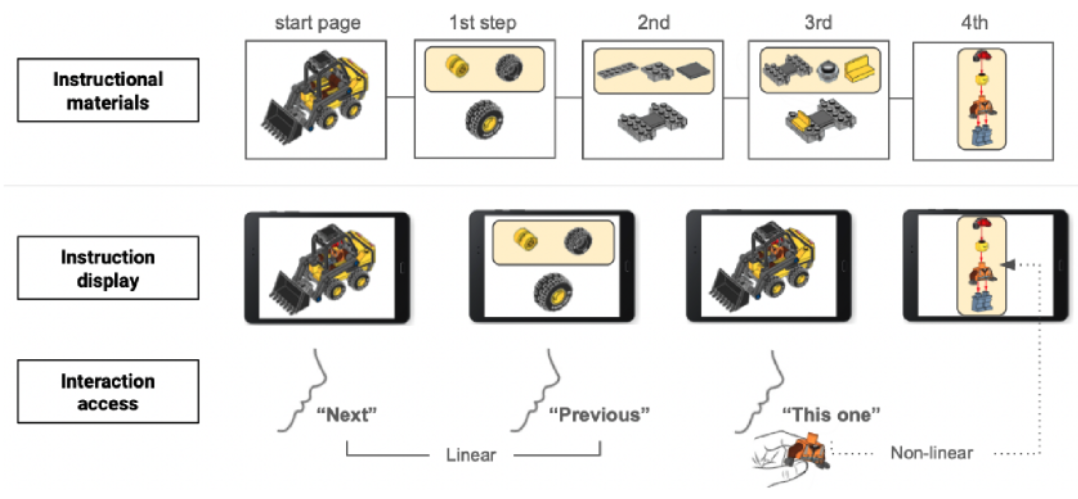}
    \caption{In the context of the toy-block assembly task, with SelfIE assembly instructions, assemblers can flexibly navigate instructions either linearly (by uttering ``next'' or ``previous'') or non-linearly (by uttering ``this one'' and showing toy block(s) towards the front-located camera). After each utterance, the corresponding step would be displayed on the screen as requested by the assembler.}
    \label{fig:Directed-Fig1}
\end{figure}

At the onset of our design process, we analysed existing literature on assembly instruction designs to identify critical factors for implementing SelfIE instructions within the toy-block assembly context. Our review emphasised the importance of minimising workflow interruptions \cite{whitlock2019authar}. Additionally, recognising the essential role of hand-eye coordination in assembly tasks \cite{feng2022seeing}, we identified the potential of leveraging alternative bodily functions to enhance user interactions with instructions for minimal disruption, such as voice commands \cite{parush2005speech} or mid-air gestures \cite{aigner2012understanding}.

Building on these findings, we initially designed three access operations for assemblers using a requester-responder interaction model. In this model, assemblers are conceptualised as requesters—not merely followers—of assistance, while the instructions act as responders, providing help as requested. The operations are designed to facilitate interactions where assemblers seek assistance through voice commands and mid-air gestures as needed to progress effectively. Fig. \ref{fig:Directed-Fig1} illustrates the initial design. 

\subsection{Design Iterations (1st \& 2nd)}
\label{SELFIE:DESIGN_ITERACTIONS}
We refined our initial design through two iterations using a Wizard of Oz (WoZ) protocol \cite{dow2005wizard} to optimize the balance between linear and non-linear access operations and ensure seamless transitions between them. The WoZ protocol allowed us to iterate on our design without the constraints of technical limitations. We adjusted the design after each round based on participant feedback and our observations. As an example of toy-block assembly tasks, we selected LEGO® due to its representative characteristic features of such tasks \cite{funk2015benchmark}.

\subsubsection*{Participants}
For the first iteration, we recruited a small sample of participants from our university community (6 participants with gender balance, age M = 25.2, SD = 8.4) to assess our design initially . In the second iteration, we expanded our participant pool to include a new and larger sample from the same community (12 participants with gender balance, age M = 27.1, SD = 4.27) to assess our design more generally.

\subsubsection*{Apparatus}
\label{SelfIE:Design_Appratus}
The WoZ experimental setup includes a MacBook laptop (13.3-inch, Apple M1 chipset, 2023 edition) and an iPad Pro tablet (12.9-inch, resolution: 2048x2732, 4th generation). We connected the tablet to the laptop as a second display \cite{AppleSecondDisplay} over the local Wi-Fi network, positioning the laptop to face the simulator and the tablet to face the participant. During the experiment, one of the experimenters took the role of the simulator, situated behind the participant, and remotely updated the displayed instructional materials on the tablet via the connected desktop according to the participant's multi-modal requests. Besides the WoZ protocol setting, we prepared three assembly models as depicted in Fig. \ref{fig:Directed-Fig2} for the training and actual test sessions. We also generated the digital instructions for each model from its original paper-based instructions provided by LEGO®.

\subsubsection*{Task Procedure}
\begin{figure}[hbt!]
    \centering
    \includegraphics[width=0.8\textwidth]{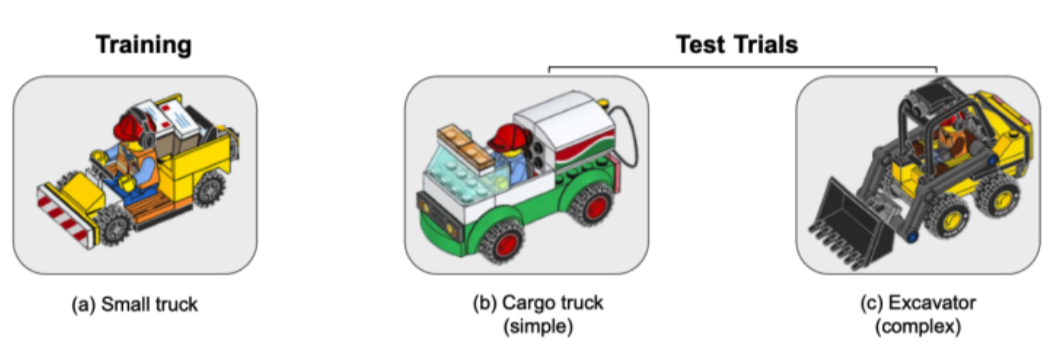}
    \caption{For our design iterations, we prepared three LEGO® vehicle models of different complexity (from left to right: the training model consisting of 24 pieces; the simple and complex models consisting of 48 and 84 pieces, respectively, for test trials to test our design in a different context robustly). To vary the complexity in the three models, we considered three factors introduced in \cite{richardson2004identifying}: the same structure among the three models but different numbers of assembly components and sub-parts.}
    \label{fig:Directed-Fig2}
\end{figure}

Before testing, we trained participants on the three access operations of SelfIE assembly instructions by having them assemble a small truck (Fig. $\ref{fig:Directed-Fig2}_a$). Following this training, participants undertook two test trials where they assembled two additional models (Fig. $\ref{fig:Directed-Fig2}_b$ and $\ref{fig:Directed-Fig2}_c$) in random order, using the SelfIE instructions. All sessions were recorded for analysis and conducted in an open-form setting without stringent requirements such as speed or accuracy. After completing the trials, we conducted open-ended interviews to gather participants’ experiences and feedback for design refinement.

\subsubsection*{1st Iteration}
During the first design iteration, our primary aim was to gather feedback on the flexible access of SelfIE assembly instructions and observe how the three operations were employed during assembly to assess our initial design and identify areas for improvement.

\leftskip 15pt
\rightskip 0pt
\vspace{10pt}
\setlength{\parindent}{0pt} \textbf{\textit{Findings and Design Refinements}}: All participants positively responded to the interaction design for using SelfIE assembly instructions. Notably, participants favoured the situational non-linear access operation for its novelty in design and ease of use, finding it as convenient as the other two linear access operations. Thereby, they attempted to employ it during the training and test sessions as needed alongside the linear access operations, leading to the identification of two issues with our initial design.

\begin{itemize}
    \item There were instances where the simulator failed to correlate the displayed assembly block with the corresponding instructional step, often due to participants' divergent assembly paths or unrecognised errors, resulting in no system response to them. When this occurred, Another experimenter had to intervene, advising participants to continue. In post-interviews, P1-2 and P5 expressed confusion over this inconsistent system behaviour. Specifically, P1 mentioned, ``Sometimes, I could proceed to the next step by showing my assembly [chunk], instead of saying `next'. But sometimes, it didn’t work for unknown reasons.'' P5 similarly said, ``Some assembly chunks worked for the jumping operation, but some didn’t. Still, I don’t understand why.'' To ensure the functional effectiveness of the situational non-linear access operation, we adjusted its mechanism by limiting it to be used with only a single LEGO® block. With this adjustment, the operation was expected to be used with a clear one-to-one mapping between the physical blocks and the digital instructional steps.
    \item There were cases in which the exact same LEGO® blocks were used in multiple instructional steps. Since handling this scenario was not considered in our initial design, the simulator randomly selected one of the multiple steps and transitioned there as a spontaneous response. In post-interviews, regarding this, P1 and P4 reported being surprised by the sudden jumping (P1, P4) and its ineffectiveness to their situations at the moment (P4). Thereby, we additionally adjusted the policy governing the non-linear access operation: if a similar scenario arises, the simulator considers the assembler's assembly context and selects the nearest step relative to the current assembly status.
\end{itemize}

\setlength{\leftskip}{0pt}
\setlength{\rightskip}{0pt}
\subsubsection*{2nd Iteration} 
\setlength{\parindent}{18pt} To confirm the effectiveness of the additional design refinements, we conducted a second round of testing using the same methodology as in the first iteration, with a new and larger sample of participants (12 participants, gender-balanced).

\leftskip 15pt
\rightskip 0pt
\vspace{10pt}
\setlength{\parindent}{0pt} \textbf{\textit{Findings and Design Refinements}}: Overall, our design refinements were effective. Participants successfully navigated the instructional materials by seamlessly blending the three access operations. However, despite these successes, two issues were further identified.

\begin{itemize}
    \item Nearly half of the participants (5/12) deviated from the prescribed assembly path, thereby using the instructional materials less frequently and making more mistakes, many of which went unrecognised. This variability in assembly approach led to challenges with the situational non-linear access operation. For instance, when P7 used the operation by presenting a LEGO® block linked to multiple steps, the simulator failed to identify an effective response despite the prior design refinements due to the participant's unidentifiable assembly path and unnoticed errors. To resolve this, we modified the non-linear access policy to allow the simulator to make heuristic decisions about which step to display in such scenarios between the linked multiple steps. This adjustment aims to provide more timely and relevant information based on the assembler's status, leveraging the simulator’s capacity to guide assembly without extensive analysis of the assembler’s exact position relative to the instructions. This efficiency was possible because the simulator memorised all steps and continuously observed the assembler's status.
    \item During post-interviews, a few participants emphasised the need to restructure instructional materials for better flexible access. They pointed out that the existing order and presentation were too reliant on linear usage assumptions, causing confusion and fragmented assembly experiences. Interestingly, all the participants recommended incorporating an exploded view \cite{Exploded-view-drawing} into the materials, which displays assembly components in a hierarchical, layer-by-layer format. One represented their reasoning for the suggestion, stating, ``So, I can access or explore each component independently via the proposed flexible access.''
\end{itemize}

\leftskip 0pt
\rightskip 0pt
\vspace{10pt}
\subsubsection*{Summary Of Design Iterations} 
\setlength{\parindent}{18pt} In summary, the initial design for assemblers' flexible access underwent two key refinements across two rounds of design iterations. The first refinement revised the policy for the situational non-linear access operation to restrict its use to a single LEGO® block, ensuring a precise 1-to-1 mapping between the physical block and the corresponding instructional step. This change aimed to balance the non-linear and linear access operations effectively. The second refinement addressed ambiguities in executing the non-linear access by allowing the system to make heuristic decisions based on the assembler's context and status, thus smoothing the operation's execution. Besides these refinements, the iterative design process revealed another potential refinement for enhancing assemblers' flexible access: redesigning instructional materials to better suit flexible navigation by removing connections between each step.
\section{Comparative Study 1: Using Instructions via Linear-Only versus Flexible Access}
\label{SELFIE:COMPARATIVE_STUDY_1}
After completing the design embodiment phase, we shifted our focus to evaluating the usability and user experience of SelfIE assembly instructions. Our design, which integrates both linear and non-linear access, was compared with traditional instructions providing only linear access. This comparison aimed to empirically validate our design approach and establish a baseline for understanding the differences in using instructions with flexible access.

\vspace{0.3cm}

\noindent This study obtained approval from the university's ethics committee, and our research questions follow:

\begin{enumerate}[\label={}]
    \item \researchQuestion{RQ}{1} How do users accept using instructions with flexible access? 
    \item \researchQuestion{RQ}{2} How do users use instructions with flexible access during assembly?
    \item \researchQuestion{RQ}{3} What impact do users have on their assembly from using instructions with flexible access?
\end{enumerate}

\vspace{0.05cm}

\subsection{Study Design}
We employed a within-subjects design with two levels of instructional navigation methods as the independent variable: linear-only and flexible access (referred to as conventional and SelfIE instructions, respectively). The order of instructions was counterbalanced, and each was randomly matched with one of two assembly models of similar complexity. Our dependent variables included objective measures (e.g., accuracy and time) and subjective aspects of user experience (e.g., perception) when using instructions with different navigation methods. To ensure a fair comparison, we controlled participants to navigate conventional instructions in a voice-based manner equivalent to the linear access operations of SelfIE instructions. This was done because speech-based interaction is often recognised as advantageous for multitasking situations involving hands-busy and eyes-busy scenarios, such as assembly tasks \cite{parush2005speech}. In addition to the standardisation of the linear navigation, we took measures to minimise potential effects on our measurements, as described in the following section.

\subsection{Apparatus}
Our comparison utilised the same WoZ protocol employed in our design embodiment. Therefore, the setup and apparatus for our comparative study were consistent with the description provided in Section \ref{SelfIE:Design_Appratus}. Since the WoZ protocol required the experimenter to simulate the system, we made two tweaks to the simulation to minimise any potential effects on measuring task completion time. First, we prepared the assembly models using unique blocks, ensuring each block appeared only once in the instructional materials. This approach created a 1-to-1 mapping between the blocks and the instructions, minimising ambiguity during the assembly process. Second, the simulator memorised the complete mapping between each block and its corresponding step for each model to be assembled. Additionally, a cheat sheet was given to the simulator so as to refer to it in case of a failed recall.

In preparing the assembly models and their instructional materials for our comparison, we meticulously crafted them to ensure fairness in assembly complexity. Details of this preparation can be found in Appendix \ref{SelfIE:assembly_material_creation}. Note that for this study, we opted for mega-bloks instead of LEGO® due to their greater flexibility in customising abstract models and controlling complexity levels. Additionally, to simplify comparisons, we excluded all advanced functionalities helping instructional navigation, such as zooming in and out.

\subsection{Participants}
The recruitment process involved 24 participants (11 males, M=22.6, SD=2.3) from the university community. None of the participants had colour blindness. Each participant received approximately 10 USD for every hour of participation.
 
\subsection{Procedure \& Task}
Participants began the experiment by providing consent and completing a preliminary survey on their prior toy block assembly experiences (e.g., LEGO®) and preferred assembly styles. The experimenter then gave an overview of the study and introduced two instruction methods: linear-only and flexible access (referred to as conventional and SelfIE instructions). Participants had opportunities to train with both methods to ensure equal familiarity. They were asked to practice all three access operations until fully comfortable and encouraged to seek clarification if needed. Following the training, participants underwent two test trials, each focusing on either conventional or SelfIE instructions. Before each trial, they were informed they could use instructional materials as they wished, including reading and scanning them beforehand. They were instructed to assemble the models in an open-ended form, without specific time or accuracy guidelines. Participants initiated each trial by stating ``start,'' after which the experimenter provided the model's blocks. Upon completion, participants signalled ``finish'' to the experimenter. Afterward, participants rated their experimental experience and the use of SelfIE instructions through a questionnaire and engaged in a semi-structured interview. The entire process was counterbalanced and recorded for analysis.

\subsection{Measurements}
Our measurements included both objective and subjective measures, outlined as follows:

\subsubsection*{Objective Measures}
\begin{itemize}
    \item \textbf{\textit{Accuracy}}: was measured as the percentage of correctly assembled blocks for the given assembly model. The policy for counting the correctly assembled blocks was that if each block's bottom part was adequately aligned and fastened with other blocks according to the instructional materials, it was counted as correctly assembled.
    \item \textbf{\textit{Task Completion Time}}: was measured from when the participant said ``start" to when s/he said ``finish". We excluded any time delay resulting from the experimenter's mistake in simulating the experiment or pouring the blocks in front of the participant via post-video analysis.
    \item \textbf{\textit{Voice Commands}}: were counted as the total number of ``previous", ``next", and ``this one" uttered by participants during assembly. Note that for conventional instructions, participants could only utter ``previous" and ``next" and not ``this one". Additionally, all voice command utterances were encoded by two independent coders through analysing the recorded trial videos. Initially, the coders annotated each utterance with its corresponding timestamp. Sequentially, they provided situational descriptions for the participant's assembly context, along with their interpretation of his/her intention behind each utterance. Following this encoding, the primary researcher reviewed the initial analyses alongside the original videos. Any discrepancies in the encoding were resolved through discussion with the two coders. After the initial review, the group meticulously re-examined the final analyses, engaging in discussions until all concerns were addressed. The analyses were then employed to identify and categorise patterns and purposes using the situational nonlinear access operation. Thematic analysis \cite{clarke2021thematic} was employed for this purpose.
\end{itemize}

\subsubsection*{Subjective Measures}
\begin{itemize}
    \item \textbf{\textit{Prior Experience Level}}: was examined in terms of two dimensions (frequency and confidence levels in performing toy-block assembly) using a 7-point Likert scale.
    \item \textbf{\textit{Preferred Instructional usage}}: was gathered by asking which one is preferred between reading instructions before assembling (instruction-first) or vice versa (task-first), as studied in prior research \cite{ganier2004factors,eiriksdottir2008people}.
    \item \textbf{\textit{User Preference}}: was measured by the participant's binary choice of either using SelfIE or conventional instructions for the same assembly tasks in the future.
    \item \textbf{\textit{Cognitive Load}}: was evaluated by using Raw NASA TLX (RTLX \cite{hart2006nasa}).
    \item \textbf{\textit{WoZ Simulation Fidelity}}: was gauged by asking how closely the system was simulated to an existing system, using a 7-point Likert scale.
    \item \textbf{\textit{Assembly Complexity Similarity}}: was assessed by asking about the perceived similarity in complexity between the two assembly models used in the test trials, using a 7-point Likert scale.
    \item \textbf{\textit{Semi-structured Interviews}}: were conducted based on filled questionnaires and trial observations. The experimenter asked participants to elaborate on their experiences based on their choices or answers in the questionnaire. Before starting the interviews, participants were allowed to review their questionnaires and change their minds by reflecting on their latest and most accurate ideas. All interviews were recorded and transcribed for further thematic analysis \cite{clarke2021thematic}.
\end{itemize}

\subsection{Data Analysis}
We used Shapiro-Wilk to test if our data is of normal distribution. Then, we used either a parametric test (paired-samples t-test) denoted by ``P" or a non-parametric test (Wilcoxon signed-rank test) denoted by ``NP". For the analysis of Likert scale data, we utilised mean scores. In all data analyses, we accepted the statistical significance at p < 0.05 and rounded up the analysis results to the second digit after the decimal point if there was no particular reason.

\subsection{Results}
Overall, participants rated our WoZ simulation to have performed quite close to a system existing in reality (M=5.67, 7: very realistic). They also perceived that the assembly models they assembled during test trials were of similar complexity (M=5.32: 7: very similar). Of the 24 participants, three (P9, P11, P19) did not use the situational nonlinear access operation while using SelfIE instructions. It means that these participants completed the given assembly model by using only the ``previous" or ``next" utterances, similar to their usage of conventional instructions. Because these participants lacked the opportunity to compare SelfIE assembly instructions with conventional instructions, we separated their data from the primary analysis (of 21 participants) and interviewed them about why they did not use the situational nonlinear access operation during their assembly. Additionally, we categorised the 24 participants into three groups based on their self-reported confidence and frequency in assembling toy blocks: High Prior Experience Level (P2, P3, P5, P7-8, P12, P21, P23), Medium (P1, P4, P9, P11, P14, P18, P20, P22), and Low (P6, P10, P13, P15-17, P19, P24).

\vspace{0.3cm}
The following results are organised in the order of measurements that are closely relevant to answering each research question. Additionally, user recommendations for further enhancements are included as a separate section, based on insights provided by participants during their post-interviews.

\vspace{0.3cm}
\setlength{\parindent}{0pt} \textbf{\researchQuestion{RQ}{1}: How do users accept using instructions with flexible access?}

\setlength{\parindent}{18pt} 
\subsubsection*{Cognitive Load} 
The RTLX analysis revealed that participants perceived higher cognitive load (P: t = -2.682, p = .014) when using SelfIE instructions (M = 66.19, SD = 19.16) compared to conventional instructions (M = 51.43, SD =19.16). In post-interviews, two participants who reported experiencing high cognitive load with SelfIE instructions explained that they explored instructional materials using flexible access to achieve their goals, resulting in increased cognitive load compared to conventional instructions. Specifically, P1 mentioned, ``I explored the necessary parts of instructional materials to understand the model's structure better, which required more brain energy.'' Additionally, P14 stated, ``I tried to assemble the model in my own way, so I explored necessary parts of the instructional materials as needed and relied on my memory to connect the scattered information during assembly.'' Besides the two participants, others who aimed to complete the task with minimum time and effort explained that they sought the most efficient assembly path when using SelfIE instructions, rather than solely relying on the order of the instructional materials. For instance, during assembly, these participants constantly questioned which remaining subpart would be the easiest to assemble (P2, P5-7, P16-17) or how they could increase efficiency in navigating the instructional materials (P5-7, P14, P16, P22), leading to having them think more during their assembly.

However, interestingly, 62\% (13/21) of participants viewed the increased cognitive load positively. For instance, P2 mentioned, ``Even though [using] SelfIE [instructions] got me to spend more time and mental effort, I am still fine with that because that is my preferred way of doing the assembly.'' Additionally, P14 deemed the increased cognitive load as ``a reasonable cost'' for the flexibility in navigation offered by the instructions. Similarly, P1 shared this sentiment by stating, ``Such additional [cognitive] effort is acceptable to understand better the model to be assembled.''

\subsubsection*{Engagement Level}
The UES analysis showed that participants felt more engaged during assembly using SelfIE than conventional instructions. First, participants found using SelfIE instructions (M=4.09, SD=0.63) to be more rewarding (NP: Z = -2.083, p = .039) than conventional instructions (M = 3.69, SD = 0.68). For this result, participants shared several sentiments: SelfIE instructions provided a ``new" (P7) and ``interesting" (P2, P8, P16) experience; its flexible access design was ``intuitive" (P7-8, P22) to use; especially, the situational non-linear access operation made it ``easier and faster" (P10, P14, P17, P22-23) to access any step of the instruction. Second, participants found using SelfIE instructions (M = 3.99; SD = 0.69) to offer more focused attention (NP: Z = -2.07, p = .041) than conventional instructions (M = 3.86; SD = 0.63) in that they could decide where to access (P1-2, P4, P8, P16) and how to perform during the task with the ease of exploring instructional materials (P1, P13-14).

\subsubsection*{User Preference}
Our questionnaire analysis indicated that 71.43\% of participants (15/21) chose to use SelfIE instructions in the future if they were to redo the same assembly tasks. Those who chose SelfIE instructions were evenly distributed among the three experience level bands (High: 5, Medium: 5, Low: 5). Among the 15 participants who chose to reuse SelfIE instructions, 40\% of them (6/15) preferred the task-first manner of using instructions, while 60\% of them (9/15) preferred the instruction-first manner. 

The post-interviews revealed that participants' choices were influenced by their personal viewpoints on using the situational nonlinear access operation, even though it was offered as an optional feature. For instance, those who opted for conventional instructions believed that following the sequential order of instructional materials step-by-step might be the easiest and safest way to complete the assembly models. Consequently, they expressed concerns that using the nonlinear access operation could lead to ``confusion'' (P5, P20-21, P24) or ``inefficiency'' (P15, P18, P20) during their assembly process. These perspectives were particularly pronounced among the three participants who did not utilise the situational non-linear access operation (P9, P11, and P19). For instance, P11 stated, ``There was no reason for me to use the jump operation during assembly.'' P19 expressed, ``I didn't want to take the risk of experiencing confusion with jumping here and there.'' In contrast, the majority who chose SelfIE instructions viewed the use of the nonlinear access operation positively and prioritised its benefits, such as ``efficient'' information retrieval (P1-2, P4, P6-8, P14, P16, P22-23) and ``flexible'' assembly order (P2-4, P6-8, P13-14, P17). One participant summed up the diverse perspectives by stating, ``There must be people who like to follow instructions step-by-step, but for me, the flexibility [offered by the situational nonlinear access operation] is very useful because I tend to prefer wandering off and not looking at the instructions in many real-life situations'' (P14).

\vspace{0.5cm}
\noindent \textit{\textbf{Answering \researchQuestion{RQ}{1}}}: The majority of participants positively accepted using instructions with flexible access, as evidenced by the analysed user preference (71.43\% preferring SelfIE assembly instructions). Notably, these preferences emerged consistently across diverse participants, irrespective of their prior experience levels or usually preferred styles, unless the assembler was not determined to rely on instructional materials to complete the assigned task. For those who favoured SelfIE assembly instructions, the easy-to-use non-linear access operation was effective, as its use helped them to reduce the time and effort required to locate situationally necessary information during assembly. In addition to the ease of information retrieval, the seamless blending of non-linear access with other linear accesses was appreciated, as it allowed them to tailor their instructional navigation according to their specific needs and situations during assembly. These two merits of using SelfIE assembly instructions led them to perceive significantly enhanced engagement while assembling the models, consequently leading them to view the experience as novel and intriguing, despite the increased cognitive load associated with flexible instructional navigation.

\vspace{0.3cm}
\noindent \textbf{\researchQuestion{RQ}{2} How do users use instructions with flexible access during assembly?}

\begin{table}[ht]
\tiny
\centering
\resizebox{13cm}{!}{
    \begin{tabular}[t]{rcccccccc}
        \toprule[0.5pt]
        & \multicolumn{8}{c}{\bf The total number of voice command utterances} \\
        \cmidrule[0.1pt]{2-9}
        \bf Type & 0-9 & 10-19 & 20-29 & 30-39 & 40-49 & 50-59 & 60-69 & $\leq$ 70 \\
        \midrule
        \bf SelfIE&2&1&2&6&3&6&1&--\\
        \bf Conventional&--&--&11&7&1&--&1&1\\
        \bottomrule[0.5pt]
\end{tabular}}
\end{table}
\begin{figure}
    \centering
    \includegraphics[width=0.8\textwidth]{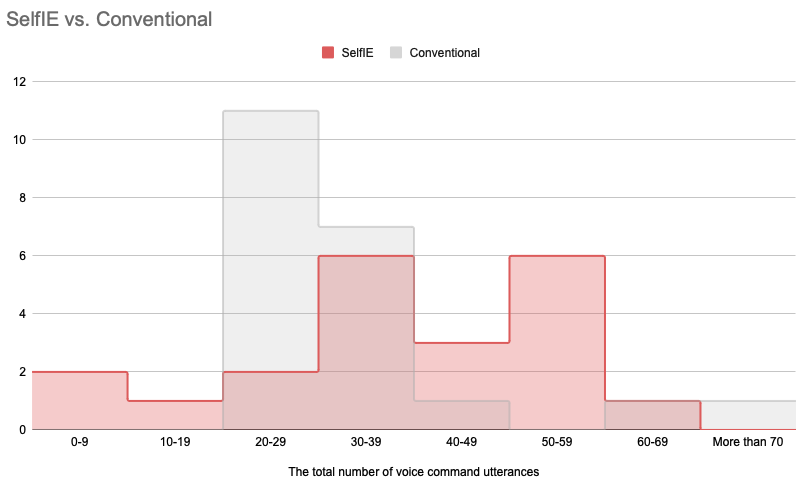}
    \caption{The provided table displays the distribution of participants across SelfIE and conventional instructions based on their total voice command utterances while assembling models of similar complexity, each consisting of 28 instructional steps. The following stepped area chart visualises this distribution, highlighting the variation across the two types of instructions. If participants carried out their assembly tasks almost equally across both types of instructions, the stepped area chart would appear almost overlapped.}
    \label{fig:Directed-Fig6}
\end{figure}

\subsubsection*{Command utterances}
This study employed two assembly models, each consisting of 28 steps. On average, while using SelfIE instruction, participants uttered the ``this one'' command (i.e., situational nonlinear access) 5.38 times (Min = 1, Max = 17, SD = 4.12) and the ``previous'' or ``next'' command (i.e., conventional linear access) 34.89 times (Min = 8, Max = 68, SD = 15.52). Our thematic analysis of these utterances revealed three primary strategies employed by participants when utilising the situational non-linear access operation, as delineated below:
 
\begin{itemize}
    \item \textbf{\textit{Selective-skipping strategy (22 out of the total 109 utterances = 20.18\%)}}: In this strategy, participants employed the non-linear access operation to expedite their instructional navigation by skipping unnecessary steps. Specifically, they scanned the current step of the instructional materials and then figured out the block or sub-part they needed to investigate further. After this identification, they jumped to the desired instructional step by locating and picking the physical block, which enabled them to do so. This strategy was used by 11 participants (P1-5, P7-8, P13-14, P16, and P20, 11/15 = 73.3\%), with nearly half of the utterances (10/22 = 45.45\%) occurring while viewing the assembly preview pages, which displayed the entire assembled model or its sub-parts before detailing each step.
    \item \textbf{\textit{Debugging strategy (23/109 = 23.1\%)}}: In this strategy, participants aimed to identify and rectify mistakes they made during assembly swiftly. Mistakes, such as missing blocks or incorrect fastenings, occurred frequently, more than we expected, with both conventional and SelfIE instructions, sometimes without the assembler's awareness. Seven participants (P1-2, P7, P12, P16, P21-23, 7/15 = 46.7\%) employed this strategy to address problematic or suspicious blocks. Instead of linearly searching for the errors within the instructional materials, they utilised the non-linear access operation to navigate back and forth between the erroneous blocks and the instructional materials, facilitating a more efficient identification of mistakes.
    \item \textbf{\textit{Block-scanning strategy (64/109 = 58.72\%)}}: 14 participants (P1-3, P5-8, P12-14, P16-17, P20, P24, 14/15= 93.3\%) employed the situational non-linear access operation in their task-first assembly approach. In this strategy, they first scanned the physical blocks placed on the table and selected where to begin by choosing one of them. Then, they referred to its corresponding instructional steps via the non-linear access operation. Unlike the selective-skipping strategy, participants' block selection was initiated from the block's physical factors, such as proximity (e.g., the block's location on the table) or attractiveness (e.g., colour or shape). In post-interviews conducted after our video and thematic analysis, P16 delineated this strategy by stating, ``I wanted to proceed with my assembly from bottom to top. So, firstly, I searched for a large and flat block, assuming that it would be a part of the model's foundation, and I began my assembly from the block.''
\end{itemize}

\begin{figure}[hbt!]
    \centering
    \includegraphics[width=1\textwidth]{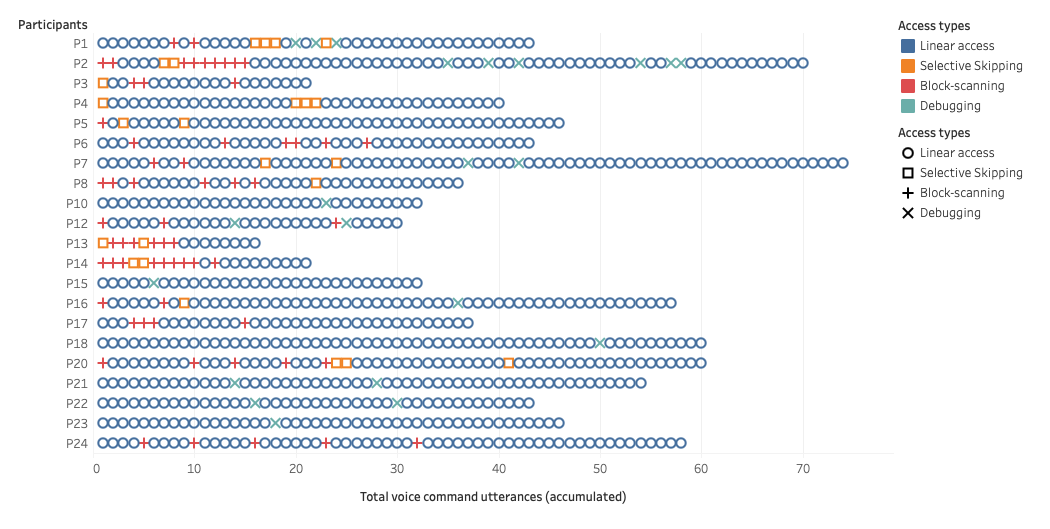}
    \caption{This chart visualises participants' diverse instructional navigation paths by mapping between voice commands uttered by each participant and access operations presenting in either linear in blue/circle or nonlinear. The operations for situational nonlinear access were further decomposed into the three identified strategies: selective-skipping (in orange/square), block-scanning (in red/plus), and debugging (in cyan/ex).} 
    \label{fig:Directed-Fig15}
\end{figure}

\subsubsection*{Observed assembly process}
Beyond categorising the non-linear access strategies, we analysed participants' overall assembly process to uncover any similarities or differences in participants' instructional usage with flexible access by comparing our qualitative and quantitative data to those of previous studies.

\begin{enumerate}
    \item As observed by prior research works \cite{fu2006suboptimal}, none of our participants read the instructions thoroughly before assembling despite the experimenter's announcement that they could do so. They adopted a mixed and simultaneous interaction between reading and assembling, and everyone started with reading the overview page.
    \item In our primary analysis of 21 participants, 15 identified themselves as instruction-first, while the remaining 6 identified themselves as task-first. In addition, their assembly experience level was almost evenly distributed into three categories (High: 8; Medium: 6; Low: 7). According to our observations, participants' instructional usages did not strictly adhere to their self-identified task or learning strategies. Their self-reported prior experience levels also had little exploratory power. These observations aligned with the findings of Eiriksdottir (2008) \cite{eiriksdottir2008people}. Instead, their instructional usage varied in the unique situations they encountered during the assigned task, as visualised in Fig. \ref{fig:Directed-Fig15}.
    \item As illustrated in the table in Fig. \ref{fig:Directed-Fig6}, using instructions with flexible access led to a more varied frequency of utterances than conventional instructions. This caught our attention, particularly considering that participants could complete the assigned assembly tasks by following instructions linearly in both conditions, although they had the non-linear access operation in using SelfIE assembly instructions. The majority (19/21) of participants using conventional instructions completed the assigned tasks by uttering voice commands only between 20 and 40 times, except for P7. P7 uttered 111 times due to her difficulty in finding mistakes. In contrast, when using SelfIE instructions, 3 participants completed the task by uttering fewer commands (0-19 times) than total steps (of 28), while 10 participants uttered more than 40 times. This result suggested that using instructions with flexible access caused a shift in the participants' approach to completing their assembly tasks. In post-interviews conducted after our video analysis, participants attributed this shift to the ease of navigating instructional materials with flexible access (P1-5, P8-9, P13-14, P22-23). With flexible access, participants such as P4-5, P8, and P14 could tailor their instructional navigation and assembly process to better suit their situation at hand; participants like P1, P3, P13, and P22-23 found that doing so would be challenging and laborious when using conventional instructions due to its difficulty of navigation. Interestingly, two participants stated two contrasting reasons for the shift: P2 made it to have "more fun" during assembly, while P6 did it for "faster" task completion.
\end{enumerate}

\vspace{0.5cm}
\noindent \textit{\textbf{Answering \researchQuestion{RQ}{2}}}: Our results demonstrated that participants adapted their instructional navigation and assembly process to suit their individual needs and situational demands when using SelfIE assembly instructions, facilitated by the ease of navigating instructional materials. Participants made this adaptation based on their situation at hand during the assigned task, regardless of their self-claimed prior experience levels and preferred task or learning strategy. Specifically, the situational non-linear access operation was employed in this adaptation to expedite the navigation itself (selective-skipping), error detection and correction (debugging), or starting assembly from the physical blocks (block-scanning).

\vspace{0.5cm}
\noindent \textbf{\researchQuestion{RQ}{3} What impact do users have on their assembly from using instructions with flexible access?}

\begin{figure}[hbt!]
    \centering
    \includegraphics[width=0.9\textwidth]{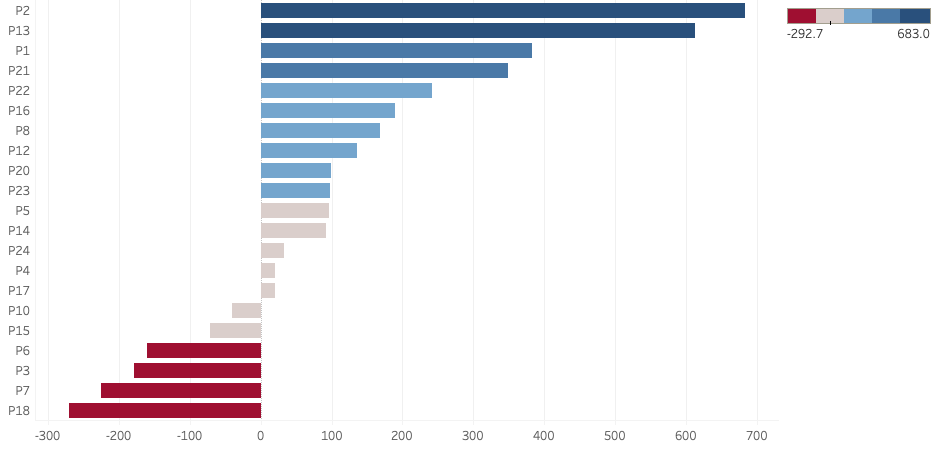}
    \caption{This figure shows the time difference between using conventional and SelfIE instructions for each participant. Note that positive values (in blue) indicate that SelfiE instructions took longer, while negative values (in red) indicate that conventional instructions took longer. The colour saturation of each bar corresponds to the magnitude of the time difference.}
    \label{fig:Directed-Fig16}
\end{figure}

\subsubsection*{Accuracy}
Out of our 21 participants, the majority (18) successfully completed the assembly regardless of the type of instructions used. Whereas, P21, when using conventional instructions, and P22, when using SelfIE instructions, achieved an accuracy of 94.4\% and 91.7\%, respectively. In addition, there was one exceptional case: P7 relinquished completing the assigned task at the end. P7 failed to recover from errors while using conventional instructions. In the post-interview, P7 stated, ``Manually checking each step [with linear access] to figure out my mistakes was so tedious and time-consuming, so I gave up [at the end]".

\subsubsection*{Time}
We excluded P7's data due to her abandonment of the task with conventional instructions, analysing the remaining 20 participants. The results showed a significant increase in task completion time (P: t = -2.31, p = .032) using SelfIE instructions (M = 591.86 sec; SD = 292.63) compared to conventional instructions (466.68 sec; SD = 292.63), indicating a 21.1\% increase. However, given the open-ended nature of the tasks, we found the need to analyse the data on a case-by-case basis. Based on this analysis, we categorised participants into three distinct groups according to the effectiveness of their use of the situational non-linear access operation in terms of task completion time.

First, P1, P2, and P13 prioritised their freedom during assembly over time. Thereby, they employed the situational non-linear access operation to maximise their freedom of task exploration and instructional navigation. As a result, they had the most significant time differences, as illustrated in Fig. \ref{fig:Directed-Fig16} at the top. P1 introduced his major as architecture in his post-interview. And, he stated, ``I explored the [model's] structure with the nonlinear access [operation] to better understand it and make my assembly plan more efficiently." P2 and P13 attempted to complete the assigned task in a task-first manner by flexibly navigating instructional materials for entertainment purposes. However, this approach led to challenges and delays in completing the task. Thereby, P13 ``changed my approach in the middle of my assembly [from task-first to instruction-first].'' Regarding this inefficiency, P2 and P13 attributed it to their inexperienced use of the nonlinear access operation. In our post-interviews, P2 elucidated this by stating, ``It seems I used the nonlinear access too much freely. My assembly order was too random.'' Consequently, P2 ``were confused with the visual difference between my assembly status and what the instructional materials showed me.'' This confusion led her to recheck the instructional materials linearly at the end, resulting in spending much more time than she anticipated. Despite the inefficient impact on their task completion time, both P2 and P16 maintained a positive view of using instructions with flexible access. They believed that with more practice, they could maximise the benefits of using instructions with flexible access in the long run.

Second, P3, P6, and P18 leveraged the situational nonlinear access operation advantageously for their task completion time. P3 and P6 marked faster task completion time when using SelfIE instructions than conventional instructions (Fig. \ref{fig:Directed-Fig16} at the bottom). According to our observation, they assembled specific sub-parts based on their choice, instead of reading and following the sequential order of instructional materials. This led them to complete the task more efficiently than following the materials, and during their assembly, they skipped certain steps by using the non-linear access operation. P18 reduced the time required to detect and correct errors by using the non-linear access operation, thereby leading to faster task completion time than conventional instructions. When detecting and correcting errors when using conventional instructions, he took more time.

Third, P3, P8-10, P12, P18, and P21 reported the cumbersome and time-consuming nature of returning to a previous step after jumping to a new one with the situational non-linear access operation. Our observation indicated that they returned to a previous step by navigating instructional materials via the linear access operations.

\vspace{0.5cm}
\noindent \textit{\textbf{Answering \researchQuestion{RQ}{3}}}: Our analysis of task performance suggested that the effect of using instructions with flexible access on performance varied based on the assembler's approach and situation taken in task completion and the individuals utilising the non-linear access operation. For example, P7 (who encountered challenges with error correction and ultimately discontinued the task) and P18 (who benefited from using the non-linear access operation in correcting errors) cases showed that using instructions with flexible access facilitated efficient debugging, thereby enhancing accuracy during assembly. However, concerning task completion time, our data indicated that the effect varied in users' ability to determine which steps to skip or consult selectively, as well as their clarity of goals during the assembly process. In particular, from the P2 and P16 cases, we observed instances where providing excessive flexibility in instruction usage could lead to inefficiencies in task performance for some users.

\subsubsection*{User Recommendations in Post-interviews}
In post-interviews, several participants provided recommendations for further improvements on the usability and experience of using SelfIE assembly instructions based on their experiences. First, P10 and P21 suggested integrating an additional ``going back'' voice command into the current design of flexible access. They expected this feature to allow for immediate backtrack, akin to executing another nonlinear jump directly to the previous step. Second, P14 shared his experience by stating, ``Repeatedly moving my arm to the camera on the table to ask about the [block's] step was tiring, and it slowed down my assembly process." So, he hoped to lay the table next to his blocks during his assembly, because ``It would make everything be right there. So, I could show the block [towards the camera] without moving around too much, and easily switch between the step and the block.'' Therein, he suggested integrating the SelfIE assembly instructions into a wearable platform, such as smart glasses. P2 and P13 shared similar experiences and ideas. Especially, P2 said, ``At the end of my assembly, I often raised blocks roughly in the air instead of showing them to the camera exactly because my arm was quite tired. If the camera is placed on my left hand, it would be great.''

\subsection{Discussion}
This section discusses the overall usability and user experience of using instructions with flexible access, along with potential enhancements based on user recommendations gathered from post-interviews.

\subsection*{Using Instructions With Flexible Access}
Our results showed that the flexible access, as proposed in the design concept of SelfIE instructions, significantly enhanced user preferences in using instructions within the toy-assembly context. As expected in our task selection, assemblers demonstrated diverse instructional approaches. Some adhered to a linear approach by regarding it as an efficient method to complete the given tasks, while others favoured a more flexible approach by prioritising their task exploration for entertainment or to discover more efficient completion paths. However, the boundary between these two groups of assemblers was not rigid. As exemplified by the case of P16, instructional approaches could shift based on individual needs and task situations. P16 initially approached instructional usage from the task-first manner by using flexible access but later switched to the instructional-first manner by primarily using linear access.

We could observe the disparity between assemblers' actual instructional usages and their self-claimed levels of prior experience or preferred styles. This suggests that individuals' unique assembly situations, such as their in-situ perceived complexity of the assigned task or intended purpose of performing it, would affect their actual instructional usages more. Particularly, the noted change in task completion behaviour when using SelfIE assembly instructions suggests that the ease of instructional navigation holds a potential as an additional influential factor in how individuals use instructions. This calls for further study.

The offered flexible access particularly benefited assemblers who prioritised task exploration. It allowed them to easily navigate and locate specific instructional steps as needed—for instance, selectively skipping unnecessary steps, directly inspecting suspicious or erroneous blocks, or referring back to instructional steps from the blocks they intended to start with. This facilitated their adaptation of instructional navigation to their individual needs and situations during assembly, in contrast to using instructions with linear-only access. As a result, they perceived their experience of using instructions as more engaging and intriguing, despite the added cognitive load, ultimately leading to a notable increase in user preference. We find this high preference meaningful for two reasons. First, rooted in our design motivation, this empirically supports the previous proposal advocating for the integration of linear and non-linear usage into instructional designs \cite{ganier2004factors,eiriksdottir2011procedural}. Second, it presents a contrasting example to prior research, which suggests that increased cognitive load when using instructions usually results in a decrease in user preference or experience \cite{ganier2004factors,redish1998minimalism,schriver1997dynamics}.

Although our new design approach led to meaningful results, its potential drawbacks also emerged. As appeared in our analysis of user preferences, the offered non-linear access operation, which enables assemblers to \textit{immediately} locate and transition to a specific step, may cause confusion and inefficiency for cautious assemblers or those who desire careful task completion. Indeed, this concern was indirectly exemplified in P2's case, whose indiscriminate use of the non-linear access operation resulted in confusion and delays in task completion. However, it is important to note that such reliant or cautious assemblers can also benefit from having the non-linear access operation, as observed in the cases of P7 and P18, where its use lowered mental barriers (P7) and saved time (P18) in error correction. Hence, we see a need and opportunity to refine the current design embodiment to embrace different types of users and their varying needs more delicately.

\subsection*{Enhanced Flexible Access Based on Cognitive Processes}
As one approach to accommodate diverse users and their needs in using instructions with flexible access, we propose sophisticating assemblers' utilisation of flexible access by providing support for their cognitive processes.

Ganier et al. \cite{ganier2004factors} proposed a mental model that elucidates users' cognitive processes in using procedural instructions during tasks. We see an opportunity to apply this model within the context of toy-block assembly, given its comprehensive description of the mental model in a general sense. This model first identifies five cognitive activities: 1) setting/holding a goal representation; 2) integrating information from the instructional materials, the task-performing situation, and the user's prior knowledge; 3) action planning/executing; 4) activity monitoring/regulation; 5) integrating into long-term memory. The user starts with the first activity to perform the assigned task (i.e., reliant or self-directed). Then, the user repeats the second and third activities until the user's initial set goal is achieved while doing the fourth activity for error correction in parallel. The last activity is related to performing the task only once or repeatedly for knowledge acquisition.

To better align assemblers' flexible instructional usage with their mental model, the three identified non-linear access strategies (i.e., selective-skipping, debugging and block scanning) can be transformed into three distinct modes. Thereby, based on specific needs for flexible navigation during assembly (i.e., the 1st activity), assemblers can select one of these modes through explicit interactions. Additionally, in the selective skipping mode, an overview of the instructional steps can be presented concurrently, allowing assemblers to devise an action plan by integrating an understanding of where they are and which steps were skipped (i.e., while performing the 2nd \& 3rd activities). This may help reduce confusion while in their flexible navigation and order of assembly. Besides, in the debugging mode, it can work on a secondary screen separately because the assembler's error correction will be triggered by the parallel monitoring activity in their cognitive processes (i.e., the 4th activity). Finally, in the block-scanning mode, the overview of instructional steps can highlight the relevant steps associated with the displayed block, aligning with the mode's sub-structures. This visual aid may alleviate the mental strain for the assembler, making it easier to identify other relevant steps and blocks in the phases of information integrating and action planning (i.e., the 2nd \& 3rd activity). This is also expected to guide them to avoid haphazard block assembly, which ultimately causes difficulty in applying the instructional materials to their assembly situations. These proposed design enhancements call for further study to validate its effectiveness.

\subsection*{Enhanced Flexible Access Based on User Recommendations}
\label{SELFIE:ENHANCED_FLEXIBLE_ACCESS_BASED_ON_USER_RECOMMENDATIONS}
Based on user recommendations and our observations, we see another approach to enhance assemblers' utilisation of flexible access.

First, as recommended by P10 and P21, improving flexible access could entail implementing an ``undo'' feature to reverse the execution of the situational non-linear access operation, akin to a ``going back'' command. Such a feature would prove advantageous in two scenarios: 1) correcting user errors in executing the non-linear operation, and 2) facilitating cross-referencing between the jumped step and its preceding step. While the "undo" operation could be done by displaying any block from the preceding step after executing the situational non-linear access, our observation through video analysis indicates that users varied in recognising this possibility. 

Second, drawing from the ideas of P2 and P13-14, we recognise the potential of implementing flexible access in a wearable configuration to enhance its usability and user experience from a new perspective. The current design embodiment is configured as tabletop (i.e., tablet-based), with the camera positioned in front of the assembler on the table, displaying instructional materials on its screen. This configuration requires the assembler to repeatedly: 1) shift his/her attention from the area where s/he is assembling to the tablet's screen and camera; 2) pick up the block that s/he wants to inquire about and move it towards the camera; 3) verbalise the "this one" voice command; and 4) wait for the inquiry result to be displayed on its screen. As proposed by P2 and P14, attaching the camera to the assembler's head or non-dominant hand (i.e., an on-body camera) and displaying the inquiry result in front of their eyes could streamline the sequence of actions required to execute the situational non-linear access operation. Indeed, for similar reasons, a considerable body of research has endeavoured to incorporate the affordance of smart glasses into the design of assembly instructions and their usage to enrich the usability and user experience \cite{blattgerste2017comparing,evans2017evaluating,blattgerste2018situ}, akin to P14's recommended idea. 

In Section \ref{SelfIE:IMPLEMENTATION}, we integrated these ideas into our translation of the WoZ-based design embodiment into a working prototype. Subsequently, we conducted further design exploration of these ideas using our working prototypes in Section \ref{SELFIE:COMPARATIVE_STUDY2}.

\subsection*{Beyond toy-block assembly task}
\label{SELFIE:DISCUSSION_BEYOND_ASSEMBLY}
The SelfIE instruction design concept was conceived to accommodate prevalent non-linear instructional approaches. As its initial design embodiment, this paper explored designing and implementing ``assemblers' flexible access'' to instructions within the context of toy-block assembly tasks. We believe the concept's embodiment can extend to other procedural tasks and their instructional designs through the meticulous design of flexible access for target users tailored to their needs and contexts. 

Researchers or practitioners aiming to extend the SelfIE instruction concept to other tasks can refer to our iterative design process (Section \ref{SELFIE:DESIGN_EMBODIMENT}) and instructional creation process (Appendix \ref{SelfIE:assembly_material_creation}) to streamline their design embodiment process. In our process, we incorporated two design refinements to resolve the observed ambiguity in using the situational non-linear access operation, i.e., the restriction of using it with a single block and the experimenter's heuristic matches. Considering that this ambiguity originated from our reliance on existing instructional materials (i.e., the Lego's official instructional materials), it is advisable to consider designing instructional materials tailored to the intended design of situational non-linear access operation. The idea of adopting an exploded view in assembly, suggested by participants, exemplifies this recommendation.
\section{Translating SelfIE Assembly Instructions to Working Prototype}

In this section, we transform the WoZ-based SelfIE assembly instructions into a working prototype to evaluate its feasibility and usability as outlined in Section \ref{SELFIE:COMPARATIVE_STUDY2}. As previously mentioned in Section \ref{SELFIE:ENHANCED_FLEXIBLE_ACCESS_BASED_ON_USER_RECOMMENDATIONS}, our prototype features both a tablet-based tabletop configuration and a wearable configuration using OHMD. This setup allows us to further investigate the potential of wearable technology to enhance the usability and user experience of instructions with flexible access.

\label{SelfIE:IMPLEMENTATION}
\begin{figure}[hbt!]
    \centering
    \includegraphics[width=0.8\textwidth]{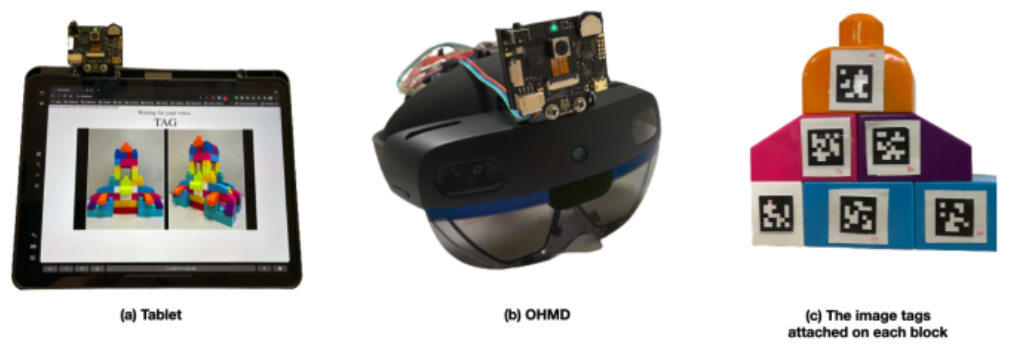}
    \caption{Prototyping: We mounted the HuskyLens Pro camera sensor \cite{husklenspro} on both (a) a tablet and (b) an OHMD to recognize (c) unique image tags assigned to each toy-assembly block. This sensor is capable of identifying the pre-assigned ID numbers of the image tags when presented by the user. Upon identification, the sensor relays these ID numbers to the connected device—either the tablet or the smart glasses—via the Wi-Fi module of the accompanying Arduino board. The transmitted ID numbers are then used to update the displayed instructions on the device.}
    \label{fig:Directed-Fig17}
\end{figure}

Our implementation of SelfIE assembly instructions comprises three key modules (Recogniser, Transmitter, and Player). Each serves a specific set of functions. The Recogniser and the Player handle the user's use of SelfIE assembly instructions, and the Transmitter bridges the Recogniser and the Player over a local wireless network connection. Fig. \ref{fig:Directed-Fig17} showcases the hardware configuration on the tablet (iPad) and the OHMD (Hololens 2). 

The three modules' incorporation allows the user to navigate instructions with flexible access via verbally commanding the player and, if necessary, showing an toy-assembly block to the recogniser, akin to the simulated SelfIE assembly instructions in our WoZ-based design exploration and embodiment.

\vspace{1em}
\noindent The following sections elaborate on each key module in detail.

\subsection{Recogniser}
The Recogniser is designed to fulfil two key functions: firstly, it detects and identifies the ID number of a tag on a toy-assembly block. Secondly, upon successful identification, it transmits this ID information to the Transmitter module. This functionality is enabled by an AI-powered HuskyLens Pro camera \cite{husklenspro} mounted on an Arduino Mega 2560 micro-controller \cite{arduino2560} equipped with an ESP8266 ESP-01 Wi-Fi chipset \cite{ESP8266}. The HuskyLens is capable of learning, detecting, identifying, and tracking image tags, which are labelled on each block as shown in Fig. \ref{fig:Directed-Fig17}c. Once an image tag’s ID is recognised, the camera processes and converts it into a digital signal on the Arduino board. This signal is then sent as a POST request \cite{Hypertext_Transfer_Protocol} to the Transmitter via Wi-Fi. The camera is configured to recognise only image tags within a specific range to reduce the likelihood of erroneous POST requests from background tag recognition.

\subsection{Transmitter}
The Transmitter relays the received tag ID information from the Recognizer to the Player. To implement this function, we developed an simple HTTP Python server \cite{HTTP_Python_server} that can handle the digital signal relay process. The server updates its internal status of tag detection if there is a POST request from the Recognizer with any identified tag ID number. And the server responds to a GET request \cite{Hypertext_Transfer_Protocol} from the Player to update its tag detection status with the relayed ID information.

\subsection{Player}
The Player component displays instructional materials and processes voice commands. We developed separate applications for the tablet and OHMD configurations. The tablet uses a JavaScript-based web application with Mozilla’s Web Speech API \cite{speechapi} to handle voice commands like ``previous'', ``next'', or ``this one''. For the OHMD, we created a Universal Windows Platform (UWP) application using Unity \footnote{\url{https://unity.com}}, which utilises Microsoft’s Windows Speech Recognition to process voice inputs. Both applications send a GET request every 0.5 seconds to the Transmitter to check for updates on user inquiries about specific blocks and update the display accordingly.

The display method varies by device. On the tablet, instructional materials appear directly on the screen, while on the OHMD, they are projected as a hologram in front of the user's sight. This hologram is dynamically adjusted to track the user’s head movements, with manual settings available for adjusting distance based on the user’s arm length and span. Unique to the OHMD, users can interact with the hologram to modify its size or position and receive audio feedback through an on-body earphone, enhancing the immersive experience.

Additionally, both player applications incorporate and support two additional voice commands compared to our prior study in Section \ref{SELFIE:COMPARATIVE_STUDY_1}: ``overview'' and ``going back''. This is a result of our reflection on the study's insights. The ``overview'' command allows the user to know which part of the instructional materials the Player displays on the screen through presenting the displayed part's step number and location over the linearly connected small box-shaped steps. The ``going back'' command enables the user to return to the previous step after using the situational nonlinear access operation.
\section{Comparative Study 2: Using SelfIE Instructions in Tabletop vs Wearable Configuration}
\label{SELFIE:COMPARATIVE_STUDY2}
After translating our WoZ-based design embodiment into working prototypes, we conducted usability assessments to pinpoint any technical hurdles or limitations that may arise during practical usage within real-world constraints. Additionally, we compared the usage of SelfIE assembly instructions between two different configurations (e.g., tabletop vs. wearable) to explore the nuanced impact of each configuration on user experience. This comparison study obtained approval from the university's ethics committee.

Given that we translated our design embodiment into the wearable configuration using the tablet-centred design without optimisation, our comparison was conducted through a qualitative investigation, focusing on the differences in user experience between the two configurations. Apart from the absence of design optimisation, the variations in performance (such as different speech recognition levels) and characteristics (like gesture-based interaction with holograms) of the implemented prototypes presented challenges for direct, objective comparison.

\subsection{Participants}
We recruited 8 participants (4 males, ave. age = 25.5, SD = 2.65) from the university community. Each received approximately 10 USD reimbursement for their 60 minutes of participation. Additional compensation was given to them proportionally to their extra participation time.

\subsection{Apparatus \& Materials}
The two working prototypes illustrated in Fig. \ref{fig:Directed-Fig17} were employed. Additionally, we created new three mega-block assembly models, one for training and the others for actual test session, including each model's instructional materials. The creation of new assembly models and their instructional materials followed the protocol we employed in our prior comparative study (Appendix \ref{SelfIE:assembly_material_creation}). The training assembly model consisted of 26 blocks, and the other two test models consisted of 83 blocks, respectively. As illustrated in Fig. \ref{fig:Directed-Fig17}c, each block has a 2cm x 2cm image tag \cite{husklenspro} on its surface. The size of image was determined based on our pilot test results, as HuskyLens Pro was able to recognise the size of image tags at the maximum distance of about 15 cm at its front under fluorescent lighting. Note that the detection length could be increased proportionally to the size of tag images. We chose the maximum distance as 15 cm by considering the users' arm span and range of motion, observed in our pilot tests.

\subsection{Procedure \& Task}
Akin to our prior comparative study, participants began by filling up the consent form, followed by training assembly trials until they reported their full readiness for the test trials. During this training session, the experimenter allowed participants to have enough time to get themselves fully familiarised with using each working prototype's features as well as accessing and reading the instructional materials with assistance. When participants reported their readiness and the experiment confirmed their familiarity, they were tasked to assemble two test models through using SelfIE assembly instructions with the tabletop and the wearable configuration of working prototype. Fig. \ref{fig:Directed-Fig17}a shows the fully assembled status of one of the test models. The order of the configuration of prototype was counterbalanced, and the test assembly models were given to participants in a random order generated by a computer. After each test trial, participants provided feedback about their experiences and the perceived usability in an open-ended interview format to the experiment. After the interview, 5 mins break was optionally given prior to continuing to the next test trial. All participants' assembly processes and interviews were video-recorded for further analysis.

\subsection{User Feedback \& Observations}
Participants' preferences for the two configurations were split in nearly half: 5/8 preferred the tabletop configuration, in contrast, 3/8 preferred the wearable configuration. The reasons for their different preferences represented trade-offs of using SelfIE assembly instructions between the two configurations, which was interesting to us. This section first discusses the results of usability assessments and moves to discussing the unveiled trade-offs in terms of each configuration's merits.

\subsubsection*{Usability}
Akin to our prior study, most participants showed their interest in using instructions via flexible access, especially, the situational non-linear access operation during interview. In average, they employed the non-linear access operation 4.64 times in the tabletop configuration and 3.81 times in the wearable configuration while assembling the test models, consisting of 28 steps. In most cases, participants could finished assembling the test models without any critical issues or errors.

However, as somewhat expected, the voice recognition and image tag detection capabilities of the AI-camera hindered participants from smoothly employing the situational non-linear access operation during their assembly process. For example, due to the prototype’s voice recognition inaccuracies, some participants had to repeat the same voice command multiple times. Others needed to re-present the block to the camera with adjustment on its displayed angle and position until it accurately recognised the block’s tag. Related to this, P2 noted the need to ``show [the block] well [towards the camera]'', which implied a challenge with the tablet’s front-facing camera. Similarly, P1 reported the cumbersome physical movements required to align with the stationary camera on the tablet.

Additionally, the maximum set distance for image tag recognition (15 cm) imposed another cognitive load on participants when triggering the non-linear access operation. Participants perceived the distance limit and figured out the fact that the camera would not recognise the displayed block's tag well out of the limited range, through their several trials and errors. This had some participants (P4, P6-7) initially check where the camera is before employing the situational non-linear access operation, even in the wearable configuration (P4), to ``avoid failing to trigger the situational non-linear access operation'' (P6). Besides, P6-7 cared about whether they would place their hand within the boundary of 15cm correctly and whether they were displaying the block at a proper angle towards the camera. Related to this, P4 and P7 stated that the limitation caused them to be ``nervous'' (P4) and ``somewhat irritated'' (P7).

As for the newly added voice commands, \textit{``overview''} and \textit{``going back''}, participants employed them during their assembly process as intended in their incorporation. For instance, P1-2 triggered the \textit{overview} command to ``monitored [their] current location over the 28 steps'' during assembly, and P6 triggered the \textit{going back} command to ``return to the previous step after using the situational non-linear access operatioin''. However, our observation for P6's case suggested that the \textit{going back} command may need a careful design to be more aligned to the command trigger's mental model: P6 uttered the \textit{going back} command repeatedly under the mistaken belief that it would act as a ``previous'' command.

These findings highlight three critical requirements for enhancing the usability of SelfIE assembly instructions within real-world constraints. First, there's a need for a more explicit and dynamic requester-and-responder interaction to facilitate situational non-linear access. For instance, users could receive visual or auditory assistance from the system when their hand enters a zone allowing accurate image tag recognition, reducing cognitive load and anxiety. Second, to address imperfect voice recognition, blocks could incorporate trackable mediums like RFID chips for automatic tracing (e.g., \cite{huang2023structuresense}). Alternatively, multi-modal input channels such as finger  \cite{sharma2021solofinger} or eyelid gestures \cite{li2020iwink} could be incorporated. Finally, voice command names should be intuitive and aligned with users' mental models, for example, using `undo'' to cancel the result of triggering the non-linear access operation instead of ``going back''.

\subsubsection*{User Merits in Wearable Configuration}
In post-interviews, P1-2 and P5 (representing 37.5\% of participants) expressed a preference for using the SelfIE assembly instructions in the wearable configuration and shared positive feedback regarding its recognised benefits. They found it easier (P1) and more intuitive (P2) to perform to show blocks towards the on-body camera attached to their head, rather than towards the front-facing camera on the tablet. P5 shared similar feedback with them. Specifically, in the wearable configuration, P1 noted that there was no additional physical movement to ``get to the [fixated] camera."; P2 found that she didn't need to ``show [the block] \textit{well} [towards the front-facing camera]." P2 depicted the difference between the two configurations by stating, ``it was like using mobile and landline phones." Besides the merit of on-body camera, participants recognised the merit of the wearable configuration from its flexibility on manipulating the location and size of the holographic instructional materials (P1, P2, P4-5). While P4 did not present any preference for the wearable configuration, he noted the merit. P1 and P2 detailed their experiences regarding the merit: they adjusted the instructional hologram to be small enough so that it did not obstruct their view, and positioned it strategically in front of their eyes. This setup enabled them to consult the hologram as needed without raising their heads during assembly,  facilitating a seamless integration of viewing and assembling activities. Overall, the feedback received was aligned with the expectations shaped by our previous study.

\subsubsection*{User Merits in Tabletop Configuration} 
In post-interviews, P3-4 and P6-8 (representing 62.5\% of participants) expressed a preference for the tabletop configuration, noting its recognised benefits. P7 and P8 reported that the wearable configuration's following instructional display caused visual disruptions during assembly, particularly when scanning blocks on the table, whereas the stationary display in the tabletop configuration did not cause the issue. Additionally, P3 described difficulty in ``finding where the hologram is in the air'' and found it ``disturbing.'' P4 and P7 echoed this sentiment; P7 particularly mentioned eye fatigue associated with locating the hologram and reading the floating materials. Regarding the eye fatigue, P7 and P8 noted that the tabletop configuration provided more comfort to eyes due to its stability (P7) and higher resolution (P8), which made reading the instructional materials easier and less tiring. The feedback received highlighted trade-offs between the tabletop and wearable configurations for using SelfIE assembly instructions, a discovery that was not recognised in our previous study.

\subsubsection*{Hybrid Approach: Integrating Merits From Both Configurations} 
Participants' feedback highlights the potential of the wearable configuration to enhance user experiences using SelfIE assembly instructions. The wearable sensors facilitated the initiation of the situational non-linear access operation by reducing arm and body movements, as suggested by participants from our previous study. Furthermore, these sensors improved the intuitiveness of the interaction for triggering the non-linear access operation by altering the display mechanism from showing blocks to a third party to presenting them directly to oneself. In addition to the merits driven from the wearable sensors, the holographic display of instructional materials offers users another flexibility in their using instructions that allows them to customise the size and placement of the displayed materials to suit their needs and specific situations during assembly, such as streamlining the interleaving between blocks and instructional materials.

However, the challenges of locating and reading instructional materials due to their movement, combined with the display's low resolution, suggest that maximising the potential may require a hybrid approach. This would integrate the benefits of both the tabletop and wearable configurations. In prior studies on Optically Head-Mounted Displays (OHMDs), Billinghurst \textit{et al.} \cite{Billinghurst1998} identified three categories based on the information’s fixed reference point: \textit{world-locked} (anchored to the world), \textit{head-locked} (anchored to the head), and \textit{body-locked} (anchored to the wearer’s body). Feedback from participants who preferred the tabletop configuration indicated that the head-locked display can obscure the location of instructional materials and interfere with viewing other visual elements like building blocks or the environment. Rzayev \textit{et al.} \cite{rzayev2020} also found a similar intrusiveness within the context of face-to-face conversations, thereby, they recommended an \textit{observer-locked} display, which positions secondary information at a fixed location relative to the conversation partner. We find this recommendation aligns with the recognised merits of the tabletop configuration by participants. Furthermore, from P7 and P8's feedback, we expect that that integrating either a \textit{world-locked} or \textit{body-locked} display with the SelfIE assembly instructions in the wearable configuration will make reading instructional materials easier for users, as it can help to minimise afterimages associated with their movements on the low-resolution display.

Hence, ideally, employing SelfIE assembly instructions in the wearable configuration would enable users to seamlessly transition among the three categories of displays—head-locked, body-locked, and world-locked—for instructional materials. For example, when the user’s posture remains relatively stationary, a world-locked display can be used to maximise stability on the display of instructional materials. However, if the user moves significantly—either relocating or changing the viewing angle for an extended period—the display can switch to a body-locked mode to place the instructional materials nearby the user’s new location or view-angle according to the user's new body location. Additionally, when the user needs to alternate between the materials and blocks, such as when searching through a pile of blocks or checking for errors, they can summon a customised display of instructional materials at a specified size and location as a head-locked mode. This hybrid approach not only compensate the shortcomings of each display mode but also allows users to tailor the display of instructional materials to their individual needs and situations during assembly. This hybrid approach compensates the shortcomings of each display mode and facilitates their effective integration, thereby enabling users to customise the instructional display according to their specific needs and situational contexts during assembly. We believe that this increased flexibility in using instructions, particularly regarding user mobility and dynamic display of instructional materials, will collectively enhance user experiences, even combined with the navigational flexibility offered by the SelfIE instruction concept. This assertion warrants further empirical studies to confirm these expectations.

\section{Limitations}
This section address several limitations associated with the works presented in this paper.

\begin{itemize}
    \item First, as discussed throughout this paper, the design of SelfIE instructions aims to enhance usability and experience of using instructions by embracing people's non-linear instructional usages into the design, rather than focusing on enhancing user performance. Therefore, applying the concept to designing instructions for procedural tasks that demand high user performance or require users to strictly follow the sequential steps might not be ideally suited. However, even in such procedural tasks, we believe that the advanced and user-friendly non-linear search operations could still contribute to user performance, as implied in our comparative study detailed in Section \ref{SELFIE:COMPARATIVE_STUDY_1}. Such feature would facilitate the user to debug his/her errors that can always occur during procedural tasks, thereby potentially increasing overall task efficiency (in terms of time) and effectiveness (in terms of accuracy).
    \item Second, researchers and designers aiming to implement flexible access in other tasks may need to develop custom interaction models and designs tailored to their target user groups. For example, when applying the concept of SelfIE instructions to cooking instructions, the design of flexible access for cooks may differ from that for assemblers due to the unique culinary environments. Unlike assemblers, cooks often have wet or unclean hands and work in noisy conditions, making request-and-responder model-based interactions impractical. Instead, it may be advantageous to configure the system to act as an observer, with cooks as practitioners. This observer-and-practitioner interaction model could enable flexible access for cooks through gaze interactions, allowing them to initiate non-linear access using gaze directions combined with voice commands or facial expressions, such as eyelid gestures \cite{li2020iwink}.
    \item Third, our comparative study outlined in Section \ref{SELFIE:COMPARATIVE_STUDY_1} deliberately excluded auxiliary functions like zooming and scrolling, as well as synergies with prior works that could enhance users' interaction with instructional materials. Therefore, the potential impact of integrating such functionalities and leveraging existing works with the SelfIE instructions concept remains to be investigated. For instance, exploring how aids like zooming in/out could facilitate users in navigating 3D/AR-based instructions linearly and non-linearly warrants further research. Thus, future studies should examine these aspects collectively to better understand their implications.
    \item Final, while our additional comparative study outlined in Section \ref{SELFIE:COMPARATIVE_STUDY2} underscores the potential of the wearable configuration to improve the user experience with SelfIE assembly instructions, further investigations are needed to solidify this finding. Since our prototyping efforts primarily focused on transforming design concepts into functional prototypes, a more detailed examination of the wearable configuration's impact on user experience is warranted through refined implementations. For instance, future iterations could actively monitor the assembler's assembly status by sensing their field of view and hand movements (e.g., \cite {huang2023structuresense}), similar to advanced approaches demonstrated in Play-and-Pause \cite{pongnumkul2011pause} or Smart makerspace \cite{knibbe2015smart}. This approach could enhance the efficiency of situational information retrieval by providing timely and relevant information tailored to the assembler's needs.
\end{itemize}
\section{Conclusion}
This paper introduces a new concept of instructional design that offers flexible linear and non-linear access, namely \textbf{Self}-\textbf{I}nitiated \textbf{E}xplorable (SelfIE) instructions. Through a WoZ protocol, we initially embodied the design concept within the context of toy-block assembly task, SelfIE assembly instructions. Subsequently, we investigate the user experience across qualitative and quantitative aspects, compared to using instructions with linear-only access. We then translated SelfIE assembly instructions into working prototypes in two device platforms (tablet and OHMD) to demonstrate its feasibility under real-world constraints and investigate its usability and the potential of the wearable configuration for an enhanced user experience with SelfIE assembly instructions.

Our study results further support the potential that enabling users to easily retrieve information from instructional materials while engaged in tasks can be a promising avenue for enhancing user experience, alongside improved comprehension and eased application. Particularly, our analysis indicates that when users could perform the information retrieval as situationally needed by blending with conventional non-linear access, user preferences were considerably increased (by 71\%) due to its ease of reflecting individual differences during tasks. This empirically supports the prior suggestion that it needs to consider people's non-linear instructional usages in the design of instructions, and our data and observations suggest that the well-designed flexible access to instructional materials can server users' diverse instructional usages, e.g., task-first, instruction-first, debugging of errors, mixed-approaches driven from the user's purpose of performing the task or depending on his/her task process. Besides, our further exploration into the wearable configuration implies that its inherent flexibility—particularly in terms of the holographic display of instructional materials’ size and location, as well as user mobility—has the potential for enhancing the user experience from a different perspective, alongside the ease of the ease of triggering situational non-linear access via on-body sensors. Building upon these findings and insights, several strategies were discussed to fully maximise the positive aspects of increasing flexibility in using instructions, such as supporting users' cognitive process or hybrid approach in the wearable configuration. These discussions call for future research to establish their effectiveness more concretely.
\section{Declaration of Generative AI and AI-assisted technologies in the writing process}
During the preparation of this work, the author(s) used ChatGPT 3.5 \footnote{http://chat.openai.com/} in order to improve and polish readability and clarity in writing. After using this tool/service, the author(s) reviewed and edited the content as needed and take(s) full responsibility for the content of the publication.

% Use the style of numbering in square brackets.
% If nothing is used, default style will be taken.
%\begin{enumerate}[a)]
%\item 
%\item 
%\item 
%\end{enumerate}  

%\begin{itemize}
%\item 
%\item 
%\item 
%\end{itemize}  

%\begin{description}
%\item[]
%\item[] 
%\item[] 
%\end{description}  

% Figure
% \begin{figure}[<options>]
% 	\centering
% 		\includegraphics[<options>]{}
% 	  \caption{}\label{fig1}
% \end{figure}

% \begin{table}[<options>]
% \begin{tabular*}{\tblwidth}{@{}LL@{}}
% \toprule
% \midrule
%  & \\
%  & \\
%  & \\
%  & \\
% \bottomrule
% \end{tabular*}
% \end{table}

% Uncomment and use as the case may be
%\begin{theorem} 
%\end{theorem}

% Uncomment and use as the case may be
%\begin{lemma} 
%\end{lemma}

\appendix

% \section{}\label{}


\bibliographystyle{cas-model2-names}

\bibliography{cas-refs}

\newpage
\appendix

\section{ASSEMBLY MODEL AND INSTRUCTIONAL MATERIALS CREATION}
\label{SelfIE:assembly_material_creation}
We prepared assembly models and their instructional materials for our studies by considering three features.

\subsection{Assembly Model Creation With Three Different Complexities of Sub-parts}
We constructed an assembly model comprising three sub-parts with differing complexities: \textit{low} complexity involved 4 blocks in 2 steps, \textit{medium} complexity consisted of 10 blocks in 5 steps including 2 asymmetric elements, and \textit{high} complexity featured 22 blocks in 13 steps with 3 asymmetric components. The model was designed to be assembled progressively, culminating in a final step that joins all three sub-parts, thereby totalling 21 steps. This design was intended to encompass a range of assembly contexts, aligning with Richardson's guidelines \cite{richardson2004identifying} for differentiating the complexities based on step count, block count, and symmetry. To validate the complexity distinctions, we conducted trials with six testers who assembled each sub-part using its specific instructions. Their completion times and perceived difficulty were statistically analysed. ANOVA and Games-Howell nonparametric post-hoc analysis (Levene test p = .003) confirmed significant variances in completion times (all p < .05), and a Kruskal-Wallis test revealed significant differences in the testers' perceptions of complexity (H (2) = 13.35, p = .001).

To tune each complexity of sub-parts, we primarily altered the number of blocks used and the number of sub-parts. In parallel, we refined the instructional materials of the three sub-parts (e.g., shown angles, colours, etc.) based on pilot testers’ feedback. This was done because the quality of instructional materials also affected their perception of assembly complexity. With the final version of instructional materials, testers completed the assembly model composed of the three sub-parts without any errors (4/6) or with a few errors (2/6). The errors were measured by the number of assembled blocks misaligned with the given instructional materials. Although the two testers made more mistakes in the increased assembly complexity, there was no significant statistical difference in testers’ accuracy between the three sub-parts. This confirms the quality of the presentation on instructional materials was improved enough.

The average completion time for the finally prepared assembly model was 542.85 seconds among 6 participants, which was close to the sum of the average total completion time for each subset (\textit{high}: 338.45 sec, \textit{medium}: 338.45 sec, \textit{low}: 44.46 sec) but somewhat longer because of the last additional step for combining the three sub-parts.

\subsection{Two similar assembly models}
For a within-subjects comparison design, we crafted a second assembly model after creating the first. To ensure similarity in complexity and assembly steps between the two models, we used the same mega-blocks from the initial model but slightly adjusted the second model's structures. Additionally, we aimed to maintain consistency in instructional materials between the two models. Six pilot testers assembled both models in a counterbalanced order, and their performance and perceptions were evaluated using the same measures employed during the creation of the first model. This process was conducted to confirm the comparability of the two models. In our comparison, the results of paired-samples t-test showed no significant difference in the testers’ average completion time (t (5) =1.096, p = 0.323) and perceived difficulty (t (5) = 0, p = 1 > .05). Both measures satisfied the normality assumption (tested by Shapiro-Wilk, all p values > .05), and all testers completed the two models without any errors.

We prepared a separate assembly model for participant training alongside the two similar assembly models. This model was created using the same protocol employed for the previous two models with 20 mega-blocks in 20 steps. The training assembly model featured three sub-parts, with one of these levels containing two sub-sub-parts.

\subsection{Instructional materials}
Based on our further literature review and participant feedback from our iterative design process (Section \ref{SELFIE:DESIGN_EMBODIMENT}), we developed instructional materials for the assembly models based on these guidelines: (1) \textit{Picture-based materials}: Using only pictures, as commonly seen in toy block instructions \cite{funk2015benchmark}, ensures consistent quality; (2) \textit{Proper orientation for full visibility}: Each step is designed to showcase all new blocks in an optimal orientation for recognising important features, shape, and colour to maximise visibility and minimise misalignment \cite{agrawala2003designing}; (3) \textit{Layer-by-Layer presentation}: Blocks needed for each layer are grouped together in each step, reflecting the natural assembly order preferred in building toy structures \cite{zhang2017component}, and helping to clearly distinguish between steps; (4) \textit{Hierarchical assembly order}: Steps are organised from simplest at the bottom to most complex at the top \cite{agrawala2003designing,schumacher2013pattern}. Preview images of fully assembled sub-parts and their components are included before relevant steps, with sub-sub-part previews positioned at the top-left of their first step, an easily appeared feature in assembly instructions \cite{mega-block-instructions}.

We separately tested the quality of the created instructional materials several times with six testers and tuned them based on their feedback to ensure there were not any confusing points.



\end{document}